\documentclass{ws-procs9x6}
\newcommand{\be}{\begin{equation}}
\newcommand{\ee}{\end{equation}}
\newcommand{\bea}{\begin{eqnarray}}
\newcommand{\eea}{\end{eqnarray}}

\begin{document}

\title{Hadron Structure on the Back of an Envelope  
}
\author{A.~W.~THOMAS, R.~D.~YOUNG}

\address{Thomas Jefferson National Accelerator Facility \\
12000 Jefferson Ave., \\ 
Newport News, VA 23185, USA\\ 
E-mail: awthomas@jlab.org}

\author{D.~B.~LEINWEBER}

\address{Special Research Center for the
         Subatomic Structure of Matter, and \\
         Department of Physics,  University of Adelaide,\\
         Adelaide, SA 5005 Australia \\
E-mail: dleinweb@physics.adelaide.edu.au}

\maketitle

\abstracts{
In order to remove a little of the mysticism surrounding the issue of 
strangeness in the nucleon, we present simple, physically transparent 
estimates of both the 
strange magnetic moment and charge radius of the proton.  Although simple, 
the estimates are in quite good agreement with sophisticated calculations 
using the latest input from lattice QCD.  
We further explore the possible size of systematic uncertainties
associated with charge symmetry violation (CSV) in the recent precise
determination of the strange magnetic moment of the
proton.  We find that CSV acts to increase the error estimate by 0.003 $\mu_N$
such that $G_M^s = -0.046 \pm 0.022\ \mu_N$.}

\section{Introduction}
The tremendous amount of experience that 
has been gained over the last 6 years, by
studying the chiral extrapolation of lattice QCD data 
as a function of quark (or pion)
mass, has led to very important insights into hadron 
structure. The two major lessons 
learned are that:
\begin{itemize}
\item The contribution of pion loops to hadron properties decreases 
very fast as the pion 
mass increases, becoming small and slowly varying for pion 
masses above about 500
MeV~\cite{Detmold:2001hq,Young:2002cj,Leinweber:1998ej}. 
As a corollary, it follows that 
the effect of kaon loops is always 
relatively small\cite{Flambaum:2004tm} -- 
an issue we shall return to soon in the 
context of strangeness form
factors. 
\item Once the pion mass is of the order of 500 MeV or higher, 
{\bf all} hadron properties are smooth, slowly varying and 
essentially behave like the constituent quark
model. The corollary to this is that if one wishes to build a 
constituent quark model of
hadron structure, this is the mass region where it has a chance 
to work\cite{Cloet:2002eg}
--  far from the region of rapidly varying non-analytic behaviour 
associated with pions near the chiral limit.
\end{itemize}
The second lesson is of particular relevance to the understanding of 
duality, because in 
this mass region ($m_\pi > 500$ MeV) the reconstruction of the valence parton
distribution functions (PDFs) shows that each 
valence quark does indeed have a most
likely momentum fraction around 1/3~\cite{Detmold:2001dv}, 
precisely as one would expect in a naive
constituent quark picture.

One of the remarkable things that became obvious from the 
beginning of these studies
is the fact that the relatively naive cloudy bag 
model\cite{Thomas:1982kv,Miller:1984em} (CBM) 
did an astonishingly good
job of describing the mass dependence of nucleon properties, 
whether it be the mass\cite{Leinweber:1999ig}, 
magnetic moments\cite{Leinweber:1998ej} 
or moments of the PDFs\cite{Detmold:2001dv}.  
This does not mean that the CBM is ideal, 
we see from the comparison against 
$G_{En}$~\cite{Glazier:2004ny,Finn:2004tu}, in particular, 
that the sharp surface of the
MIT bag, upon which the CBM was built, 
is not such a good description of the valence quark structure, 
especially in the surface region\cite{Lu:1997sd}. 
However, what {\bf does} seem to really 
describe the way hadron structure 
works is that there is a perturbative pion cloud 
around a core of confined valence 
quarks, confined in a region whose vacuum structure (the bag itself) 
differs from that of
the QCD ground state. 

The recent discovery that the chiral quark soliton model 
also yields the correct dependence of $m_N$ on $m_\pi$~\cite{Goeke:2005fs} 
(apart from the incorrect chiral
coefficient associated with the hedgehog approximation) is consistent with this
interpretation, since even though one has to work extremely 
hard to construct the change 
in the vacuum structure inside the nucleon at the microscopic level, 
in the end it looks 
like a system of bound valence quarks surrounded 
by a perturbative pion cloud. The 
consequences of the change in vacuum structure inside the hadron, in terms of a 
contribution to $\bar{d} \neq \bar{u}$ and 
$\Delta \bar{u} \neq \Delta \bar{d}$ are in
fact similar in both models\cite{Schreiber:1991tc,Dressler:1998zi}.

Further support for this idea comes from a remarkable discovery 
concerning the lattice QCD data for the nucleon and the $\Delta$ 
in both quenched (QQCD)
and full QCD (QCD). In fact, one can describe the data 
with a simple fitting function, $\alpha + \beta\, m_\pi^2$ 
plus the pion self-energy loops 
which give rise to the leading (LNA) and next-to-leading non-analytic (NLNA) 
behaviour, evaluated using a finite range regulator of 
dipole form (with common mass
parameter $\Lambda = 0.8$ GeV).  The important discovery is that
$\alpha$ and $\beta$ (for a given baryon) 
are the {\bf same} within the fitting 
errors (a few percent) in  QQCD and QCD\cite{Young:2002cj}. 
This is the case even though for the $\Delta$ 
the self-energies differ by a factor of two, 
with the N-$\Delta$ splitting being of order 500
MeV in QQCD and only 300 MeV in full QCD. 
It seems that the ``core'' or valence
structure of these key baryons, defined by the particular value of
$\Lambda = 0.8$ GeV, is essentially the same in QQCD and QCD. The 
implications of this for modeling hadron structure are yet to be fully 
investigated but once 
again a simple, perturbative treatment of the pion cloud contributions works
exceptionally well.

One of the most impressive recent achievements of 
the chiral extrapolation program has 
been the determination of an extremely precise value for 
the strangeness magnetic 
moment, $G_M^s$~\cite{Leinweber:2004tc}. 
This calculation used a combination of 
experimental data for the
octet magnetic moments, the constraints of charge 
symmetry and chiral extrapolation 
of state of the art lattice data to obtain the ratios of 
the magnetic moments of either a 
valence $u$ quark in the proton and $\Sigma^+$ or the 
valence $u$ quark in the neutron
and the $\Xi^0$. By reducing the demands on lattice QCD 
to mere ratios, it is possible to 
dramatically reduce the systematic errors. 
The result, namely $G_M^s = -0.046 \pm
0.019 \mu_N$ is an order of magnitude more 
precise than any current 
experiment~\cite{Spayde:1999qg,Beise:2004py,Maas:2004pd,Armstrong:2005hs,Aniol:2005zg}
-- a  
unique example in modern hadron physics. A similar analysis for 
the strangeness electric 
form factor, $G_E^s$, has not yet been possible, 
essentially because the experimental 
knowledge of octet baryon charge radii is nowhere near 
as precise as the knowledge of 
magnetic moments. 
However, our main purpose, to
which we turn in the next section, 
is to use what we have learnt so far about hadron 
structure to make a ``back of the envelope'' 
estimate of both the strangeness electric and
magnetic form factors.  Then in the following section, we  provide an
estimate of 
the systematic uncertainty associated with charge symmetry violation
in the recent precise determination of the strangeness
magnetic moment of the nucleon\cite{Leinweber:2004tc}.

\section{Simple Model of the Strangeness Form Factors of the Proton}

We note first that there is no known example where 
the current quark masses show up in
hadron physics undressed by non-perturbative glue. Thus the cost 
to make an $s-\bar{s}$ 
pair in the proton is of order 1.0 to 1.1 GeV 
(twice the strange constituent quark mass). 
On the other hand, 
creating the $\bar{s}$ in a kaon and the $s$ in a $\Lambda$ costs 
only 0.65 GeV. (Note that the N to $K \Sigma$ coupling is 
considerably smaller than that 
for N to $K \Lambda$ and hence in this simple discussion we ignore it.) 
On these grounds 
alone we expect the virtual transition N to $K \Lambda$ to 
dominate the production of 
strangeness in the proton. 

Next we estimate the probability for finding the $K \Lambda$ configuration. This
probability is inversely 
proportional to the excitation energy squared. We work by 
comparison with
the N $\pi$  component of the nucleon wave function, for which there is a 
vast body of 
evidence that it is about 20\%~\cite{Speth:1996pz}. Naively the 
transition N to N $\pi$ costs 140
MeV, but with additional kinetic energy this is around 600 MeV in total. 
Including similar
kinetic energy for the $K \Lambda$ component as well, it costs 
roughly twice as much as
N $\pi$. Thus the $K \Lambda$ probability is of order 5\%.

\subsection{Strangeness radius}

We consider first the strangeness radius of the proton 
based on this 5\% $K \Lambda$
probability. In the CBM the radius of a $\Lambda$ 
bag is about 1 fm, which yields a 
mean square radius for the strange quark around 0.5 fm$^2$.  
As an estimate of the range
of variation possible, we also take the bag 
radius $R = 0.8$ fm with a corresponding
mean square radius close to 0.36 fm$^2$. In order to estimate the 
contribution from the
kaon cloud, we need to realize that in almost any 
chiral quark model the peak in the
Goldstone boson wave function is at the 
confinement (bag) radius\cite{Lu:1997sd,Thomas:2005qm}. 
As long ago as 1980 
this generated enormous interest in the precise 
measurement of $G_E^n$~\cite{Thomas:1981vc}.
The meson field then decreases with a range between one over the 
energy cost of the Fock state and $1/(m_K +m_\Lambda - m_N)$.
Thus for $R=0.8$ fm we get a 
mean square radius for the $\bar{s}$ distribution of 
order 1 fm$^2$, while for $R = 1$ fm 
we get about 1.4 fm$^2$. 
Weighting the $s$ by $-1/3$ and $\bar{s}$ by +1/3, we find 
that the mean square charge radius of strange 
quarks is between $(-0.36 + 1.0)/3$ and 
$(-0.5 +1.4)/3$, that is in the range (0.21,0.30) fm$^2$, 
times the probability for finding 
the $K \Lambda$ configuration. 

To calculate $G_E^s$ at $Q^2=0.1$ GeV$^2$ = 2.5 fm$^{-2}$, 
we assume that the term 
$- Q^2  <r^2> / 6$ dominates and finally multiply by $-3$ to agree with
the usual  
convention of removing the strange quark charge. 
This yields $G_E^s \in (+0.01,+0.02)$. 
It is definitely small and definitely positive for 
the very clear physical reasons that the $K 
\Lambda$ probability is small and that the kaon cloud extends 
outside the $\Lambda$. A comparison with the currently preferred fit 
to the existing world data\cite{Armstrong:2005hs} 
reveals that this estimate is in agreement at the 1 $\sigma$ level, although it is nominally of opposite sign. 

\subsection{Strangeness magnetic moment}

Because orbital angular momentum is quantized, the contribution to the magnetic 
moment from the $\bar{s}$ in the kaon cloud  is much less model dependent. The 
Clebsch-Gordon coefficients show that in a spin-up proton the 
probability of a spin down 
(up) $\Lambda$, accompanied by a kaon with orbital 
angular momentum +1 (0), is 2/3 
(1/3). We also know the magnetic moment of the $\Lambda$ 
and that it is dominated by
the magnetic moment of the $s$ quark. Hence the total strangeness 
magnetic moment of the proton is $-3 
\times P_{K \Lambda} \times 2/3 \times (+0.6 +1/3) -3 \times P_{K \Lambda} 
\times 1/3 \times (-0.6 + 0)$, 
where the terms in brackets are, respectively, the 
magnetic moment of the spin down (up) $\Lambda$ and the 
magnetic moment of the charge 
+1/3 $\bar{s}$ quark with one unit (or zero units) of orbital angular momentum. 
The nett result, namely $G_M^s = - 0.063\ \mu_N$, is reasonably  
close to the best lattice QCD estimate noted above, that is
$G_M^s  = -0.046 \pm 0.019 \mu_N$. From the point of 
view of this ``back of the 
envelope'' estimate, the lattice result clearly 
has both a natural magnitude and sign. It is very hard to see how 
the result could change much in either magnitude or 
sign unless the physical picture presented here is totally incorrect. 
Given the remarks concerning our present understanding of 
hadron structure based on experience with the study of 
chiral extrapolation and lattice QCD data, this seems unlikely.

\section{Impact of Charge Symmetry Violation on $G_M^s$}

\begin{figure}[tbp]
\begin{center}
{\includegraphics[height=3.3cm,angle=90]{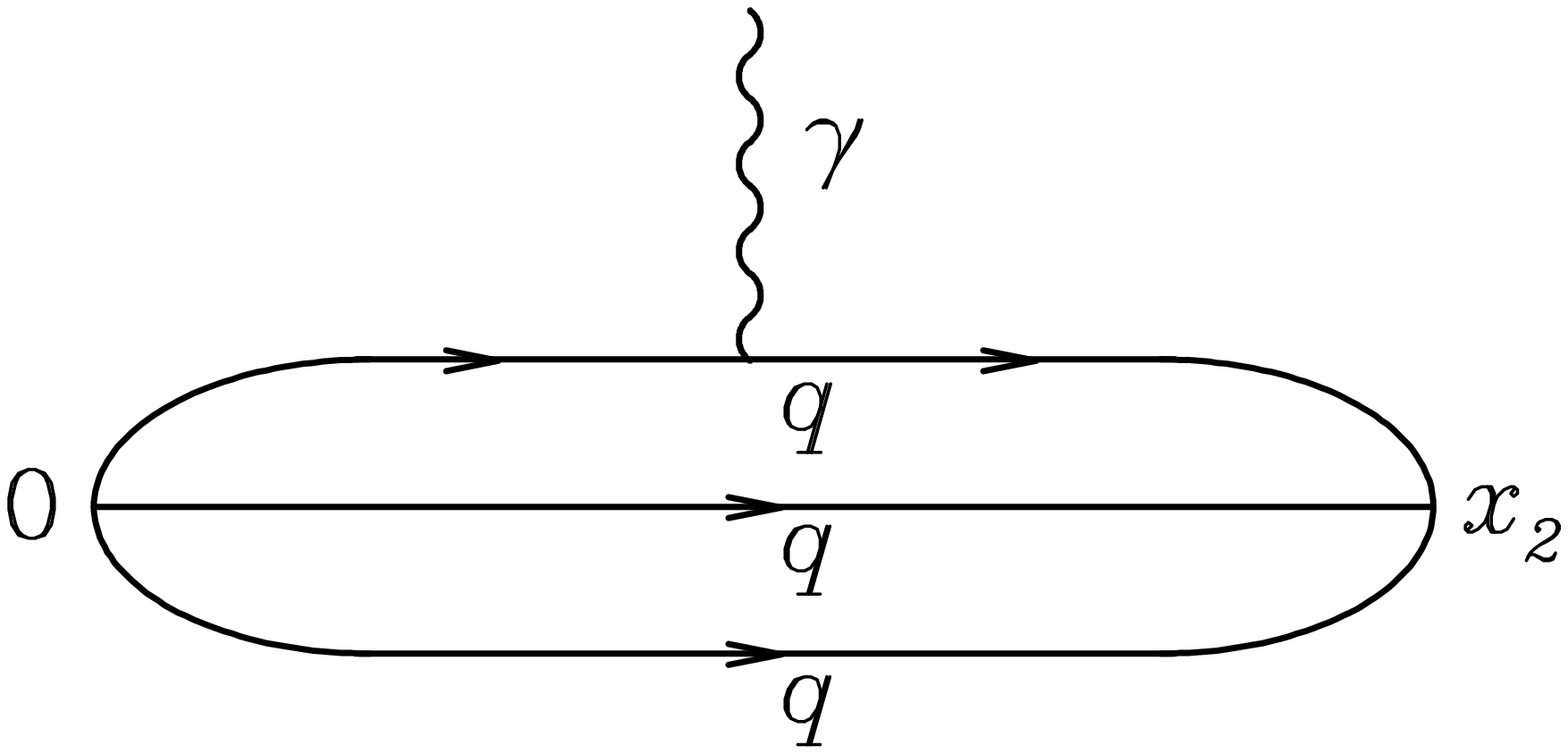} \hspace{0.8cm}
 \includegraphics[height=3.3cm,angle=90]{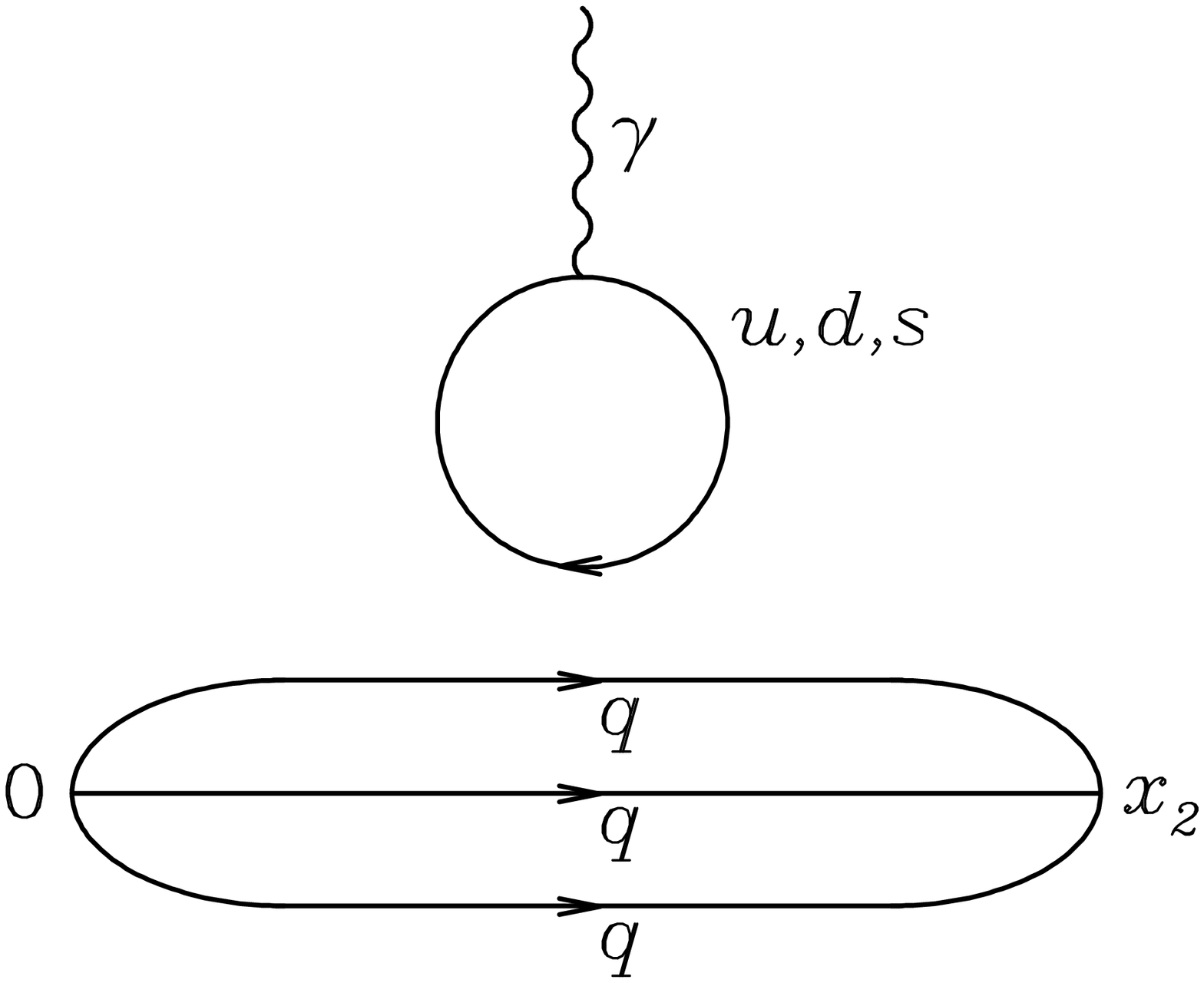}}
\end{center}
\caption{Diagrams illustrating the two topologically different
insertions of the current within the framework of lattice QCD.  
In full QCD these diagrams are dressed with
an arbitrary number of gluons and additional quark loops.
}
\label{topology}
\end{figure}

In the spirit of Refs.~\cite{Leinweber:1995ie,Leinweber:1999nf}, we
use $p, n, u^p$ etc., to denote the magnetic moment of that baryon or,
in the case of a quark, to denote the valence quark sector
contribution of that flavor to that baryon {\it if} that quark had
unit charge.  A valence quark sector contribution is depicted in the
left-hand diagram of Fig.~\ref{topology}.  We also denote the total
contribution of $u, d$ and $s$ quarks in a ``disconnected loop'' in
baryon $B$, depicted in the right-hand diagram of Fig.~\ref{topology},
as $O_B$.  By determining $O_p$, the strangeness magnetic moment of
the proton can be obtained by calculating the ratio of strange to
non-strange loop contributions.

The magnetic moments of the octet baryons satisfy:
\begin{eqnarray}
p &=& e_u\, u^p + e_d\, d^p + O_p  \, \, ; \, \, 
\hspace{10mm}
n = e_d\, d^n + e_u\, u^n + O_n  \, , \nonumber \\
\Sigma^+ &=& e_u\, u^{\Sigma^+} + e_s\, s^{\Sigma^+} + O_{\Sigma^+}
\, \, ; \, \,  
\Sigma^- = e_d\, d^{\Sigma^-} + e_s\, s^{\Sigma^-} + O_{\Sigma^-}  \,
, \nonumber  \\ 
\Xi^0 &=& e_s\, s^{\Xi^0} + e_u\, u^{\Xi^0} + O_{\Xi^0}  \, \, ; \, \,
\hspace{2.5mm}
\Xi^- = e_s\, s^{\Xi^-} + e_d\, u^{\Xi^-} + O_{\Xi^-}  \, . \nonumber  \\
\label{equalities}
\end{eqnarray}
Having removed the charge factors from the valence quark contributions
to baryon magnetic moments, it is usually at this point that charge
symmetry is invoked to provide equivalence between the
doubly-represented $u$-quark sector of the proton $u^p$ and the
doubly-represented $d$-quark sector of the neutron, $d^n$.  Similarly,
$u^{\Sigma^+}$ is taken to be equal to $d^{\Sigma^-}$, and $u^{\Xi^0}$
is taken to be equal to $d^{\Xi^-}$.  However, current quark mass
differences of a few MeV and electromagnetic effects will act to
violate these equalities.  In these cases, charge symmetry violation
(CSV) in the quark flavor being probed by the electromagnetic current
is directly related to the differences observed in baryon properties.

However, indirect environmental effects are also induced through CSV.
For example, even though it is the same strange quark that appears in
$\Sigma^+$ and $\Sigma^-$, its contributions to the baryon moment will
differ due to subtle differences in the environment of the strange
quark.  Similar environmental effects will provide subtle violations
of $s^{\Xi^0} = s^{\Xi^-}$, $O_p = O_n$, $O_{\Sigma^+} = O_{\Sigma^-}$
and $O_{\Xi^0} = O_{\Xi^-}$.

Introducing $\Delta_B$ to denote the contribution to the magnetic
moment of baryon $B$ having its origin in CSV, Eqs.~(\ref{equalities})
take the exact forms
\begin{eqnarray}
p &=& e_u\, u^p + e_d\, d^p + O_p  \, \, ; \, \, 
\hspace{10mm}
n = e_d\, u^p + e_u\, d^p + O_p - \Delta_n  \, , \nonumber \\
\Sigma^+ &=& e_u\, u^{\Sigma^+} + e_s\, s^{\Sigma^+} + O_{\Sigma^+}
\, \, ; \, \,  
\Sigma^- = e_d\, u^{\Sigma^+} + e_s\, s^{\Sigma^+} + O_{\Sigma^+} - \Delta_{\Sigma^-}  \,
, \nonumber  \\ 
\Xi^0 &=& e_s\, s^{\Xi^0} + e_u\, u^{\Xi^0} + O_{\Xi^0}  \, \, ; \, \,
\hspace{2.5mm}
\Xi^- = e_s\, s^{\Xi^0} + e_d\, u^{\Xi^0} + O_{\Xi^0} - \Delta_{\Xi^-}  \, . \nonumber  \\
\label{CSVequalities}
\end{eqnarray}
While we have elected to write the right-hand expressions of
Eqs.~(\ref{CSVequalities}) in terms of quantities in the left-hand
expressions and $\Delta_B$, we note that one could have done the
opposite and this will be important in quantifying $\Delta_B$.

The total sea-quark loop contribution to the proton magnetic moment,
$O_p$, includes sea-quark-loop contributions from $u$, $d$ and $s$
quarks (right-hand side of Fig.~\ref{topology}).  By definition
\begin{eqnarray}
O_p &=& \frac{2}{3} \,{}^{\ell}G_M^u - \frac{1}{3} \,{}^{\ell}G_M^d -
\frac{1}{3} \,{}^{\ell}G_M^s \, , \\
&=& \frac{1}{3} \,{}^{\ell}G_M^d 
- \frac{1}{3} \,{}^{\ell}G_M^s 
- \frac{2}{3} \,\Delta_{loop} \, , 
\label{Op}
\end{eqnarray}
where we have introduced
\begin{equation}
{}^{\ell}G_M^u = {}^{\ell}G_M^d - \Delta_{loop} \, ,
\end{equation}
with $\Delta_{loop}$ accounting for differences in the $u$ and $d$
sea-quark loop contributions to the proton due to direct CSV.

Introducing, in the usual fashion, the ratio of $s$- to $d$-quark loop
contributions, ${}^{\ell}R_d^s \equiv
{{}^{\ell}G_M^s}/{{}^{\ell}G_M^d}$,  Eq.~(\ref{Op}) provides
\begin{equation}
O_p = \frac{{}^{\ell}G_M^s}{3} \left ( \frac{1 -
{}^{\ell}R_d^s}{{}^{\ell}R_d^s } \right ) - \frac{2}{3}\, \Delta_{loop} \, , 
\label{OGMs}
\end{equation}
The established approach\cite{Leinweber:2004tc} centres around two
equations for the strangeness magnetic moment of the nucleon, $G_M^s$,
obtained from linear combinations of the above.  Including the
$\Delta_B$ terms to account for CSV, one has the exact relations
{\small
\begin{equation}
G_M^s = \left ( {\,{}^{\ell}R_d^s \over 1 - \,{}^{\ell}R_d^s }
\right ) \left [ 2 p + 2 \Delta_{loop} + n + \Delta_n - {u^p \over u^{\Sigma}} \left ( \Sigma^+ -
\Sigma^- - \Delta_{\Sigma^-} \right ) \right ] , 
\label{GMsSigma}
\end{equation}
\begin{equation}
G_M^s = \left ( {\,{}^{\ell}R_d^s \over 
1 - \,{}^{\ell}R_d^s } \right ) \left [
p + 2 \Delta_{loop} + 2n + 2\Delta_n - 
{u^n \over u^{\Xi}} \left ( \Xi^0 - \Xi^- -
\Delta_{\Xi^-}\right )  
 \right ] .
\label{GMsXi}
\end{equation}
}%
The ratios $u^p / u^{\Sigma}$ and $u^n / u^{\Xi}$ are ratios of
valence-quark contributions to baryon magnetic moments in full QCD as
depicted in the left-hand diagram of Fig.~\ref{topology}.  The latter
are determined by lattice QCD calculations and finite range
regularization effective field theory
techniques\cite{Leinweber:2004tc} with the results
\begin{equation}
\frac{u^p}{u^\Sigma} = 1.092\pm 0.030
\quad
\mbox{and}
\quad
\frac{u^n}{u^\Xi} = 1.254\pm 0.124 \, .
\label{uRatios}
\end{equation}
The ratio of $s$- and $d$-quark sea-quark loop contributions,
${}^{\ell}R_d^s \equiv {G_M^s}/{{}^{\ell}G_M^d}$, has been estimated
conservatively\cite{Leinweber:2004tc,Leinweber:2005jb} as $0.139 \pm
0.042$.

Tests of CSV suggest that it is typically smaller than a 1\% effect in
baryon properties.  The structure of Eqs.~(\ref{GMsSigma}) and
(\ref{GMsXi}) suggests that a good estimate of the systematic
uncertainty in $G_M^s$ would be provided by taking the CSV terms
$\Delta_B$ to represent uncertainties with a magnitude of  1\% of the
associated baryon moment.

As discussed following Eqs.~(\ref{CSVequalities}), the CSV corrections
$\Delta_{\Sigma^-}$, and $\Delta_{\Xi^-}$ could equally well have been
represented on the left-hand expressions of Eqs.~(\ref{CSVequalities})
as $\Delta_{\Sigma^+}$, and $\Delta_{\Xi^0}$.  Hence we will replace
$\Delta_{\Sigma^-}$, and $\Delta_{\Xi^-}$ in Eqs.~(\ref{GMsSigma}) and
(\ref{GMsXi}) with $\Delta_{\Sigma}$, and $\Delta_{\Xi}$ representing
the average 1\% uncertainties of the hyperon charge states.  Our focus
on $O_p$ for strangeness in the proton, does not allow a similar
symmetry for the nucleon. 

Since the underlying mechanisms giving rise to CSV, as represented
by $\Delta_B$, are different for each baryon,
the uncertainties are accumulated in quadrature.  
Focusing on Eq.~(\ref{GMsXi}) where the error is largest, 
this provides a CSV
uncertainty of 0.011 $\mu_N$ to $G_M^s$.
Given the already large error on $G_M^s = -0.046 \pm 0.019\ \mu_N$
associated with statistical, scale determination, chiral correction
and ${}^{\ell}R_d^s$ uncertainties\cite{Leinweber:2005jb}, this
additional CSV uncertainty has only a small effect on the final error
estimate.  Adding the CSV uncertainty in quadrature provides a total
uncertainty of 0.022 $\mu_N$.

\section{Concluding Remarks}

In the light of recent insights into hadron structure 
based on lattice QCD and the
associated work on chiral extrapolation using a finite range regulator, 
we have explained
how to quickly and easily estimate the strangeness electric and 
magnetic form factors of 
the proton. The resulting ranges, 
$G_E^s(0.1 {\rm GeV}^2) \in (+0.01, +0.02)$ and 
$G_M^s = - 0.063 \mu_N$ are relatively small, 
certainly challenging for our experimental 
colleagues, but consistent within 95\% CL   
with current world data. 
The latter
is also in remarkable agreement with the 
recent determination based on lattice QCD. 

We also explored the size of systematic uncertainties associated with
charge symmetry violation in the recent precise determination of the
strange magnetic moment of the proton\cite{Leinweber:2004tc}.  We
find CSV acts to increase the error estimate by 0.003 $\mu_N$ such
that $G_M^s = -0.046 \pm 0.022\ \mu_N$.  Hence even accounting for CSV
in the approach, one still has a two-sigma signal for the sign of the
strange magnetic moment of the proton.

In conclusion, this is a crucial point in the history of the study of 
hadron structure. For the
first time we have useful guidance from 
non-perturbative QCD using the methods of 
lattice QCD and chiral extrapolation. 
These rigorous calculations can be given life
through the sort of simple physical model described here, 
which nevertheless permits 
semi-quantitative calculation. At the same time we have new experimental 
capabilities to accurately measure the role of non-valence 
quarks in static properties
which can be used to test our new found theoretical advances.

\section*{Acknowledgements}
This work is supported by the Australian Research Council and by DOE
contract DE-AC05-84ER40150, under which SURA operates Jefferson
Laboratory.

\end{document}